\documentclass[%
 aip,
 jcp,%
 amsmath,amssymb,
 reprint,%
]{revtex4-1}

\usepackage{graphicx}% Include figure files
\usepackage{dcolumn}% Align table columns on decimal point
\usepackage{bm}% bold math
\usepackage{balance}

\newcommand{\vshift}{f_{\text{shift}}}

\newcommand{\xcut}{x_{\text{cut}}}
\newcommand{\vdia}{\varphi_{\text{dia}}}
\newcommand{\vsc}{\varphi_{\text{sc}}}
\newcommand{\kB}{k_{\text{B}}}
\begin{document}

%\preprint{AIP/123-QED}

\title{Phase Behavior of Materials with Isotropic Interactions Designed by Inverse Strategies to Favor Diamond and Simple Cubic Lattice Ground States}

\author{Avni Jain}
\affiliation{McKetta Department of Chemical Engineering, The University of Texas at Austin, Austin, TX 78712}

\author{Jeffrey R. Errington}
\affiliation{Department of Chemical and Biological Engineering, University of Buffalo, The State University of New York, Buffalo, New York 14260-4200}

\author{Thomas M. Truskett}
\email{truskett@che.utexas.edu}
\affiliation{McKetta Department of Chemical Engineering, The University of Texas at Austin, Austin, TX 78712}

\date{\today}

\begin{abstract}
We use molecular simulation to construct equilibrium phase diagrams for two recently introduced model materials with isotropic, soft-repulsive pair interactions designed to favor diamond and simple cubic lattice ground states, respectively, over a wide range of densities [Jain et al., Soft Matter 9 14 (2013)]. We employ free energy based Monte Carlo simulation techniques to precisely trace the inter-crystal and fluid-crystal coexistence curves. We find that both model materials display rich polymorphic phase behavior featuring stable crystals corresponding to the target ground-state structures, as well as a variety of other crystalline (e.g., hexagonal and body-centered cubic) phases and multiple reentrant melting transitions.
\end{abstract}

\pacs{Valid PACS appear here}% PACS, the Physics and Astronomy
                             % Classification Scheme.
\maketitle

Crystalline materials with simple cubic or diamond symmetries are often desired for photonic or other optical applications.\cite{PCsimplecubic,DIAphoton} Unfortunately, large-scale and inexpensive fabrication methods for such materials which allow one to prescribe both the physical dimensions and the symmetry of the periodically-replicated features have been challenging to develop.\cite{Reviewnanophotonics,PhysRepBusch2007} Holographic lithography, where a photoresist film is exposed to an optical profile with the desired symmetry created by interference of multiple coherent laser beams, represents one promising top-down approach.\cite{holographicpcscanddia} On the other hand, self assembly of micron or nanoscale particles from disordered fluid states into a targeted lattice structure is a widely studied bottom-up strategy.

\begin{figure}[t]
\centering
  \includegraphics[width=8.5cm]{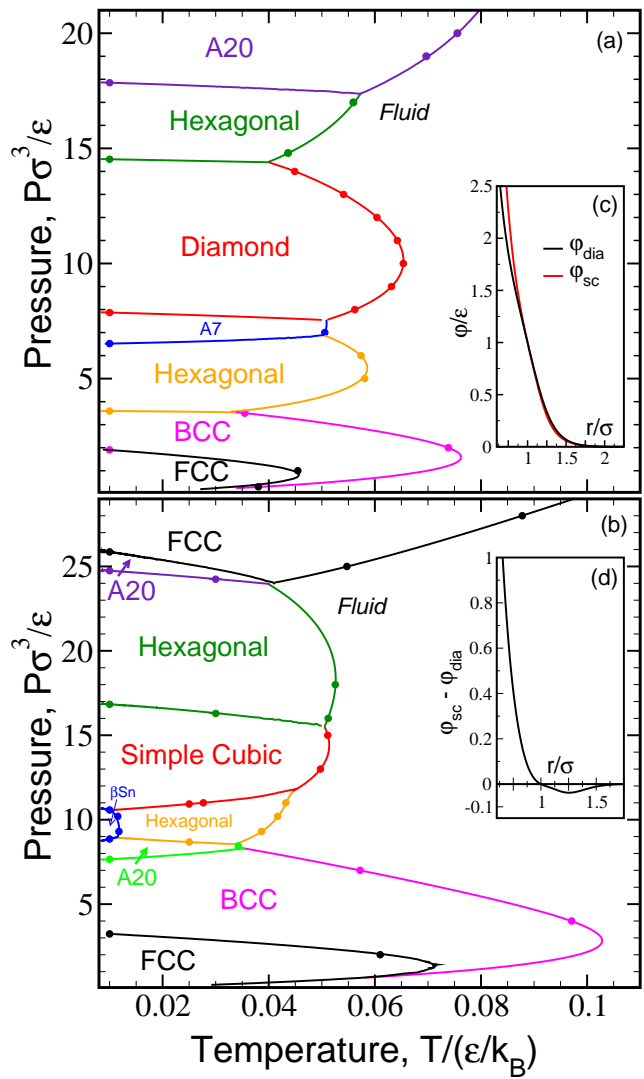}
  \caption{Pressure-temperature projection of the phase diagrams for the (a) diamond forming potential (\(\vdia\)) and (b) the simple cubic forming potential (\(\vsc\)). Points are from the reference phase coexistence simulations discussed in the text, and curves are phase coexistence traces obtained directly from the NPT-TEE MC simulations. Statistical uncertainties are smaller than the symbol size in all cases. Crystal phases are those described in the text. Inset (c) presents both potentials and inset (d) elucidates the difference (\(\vsc-\vdia\)) as a function of interparticle separation~\(r/\sigma\).}
  \label{fgr:PressurePhase}
\end{figure}

Although practical challenges remain, computer simulations and experiments have shown that self-assembly of the low-coordinated lattice structures of interest is possible by selecting particles with specific anisotropic shapes (e.g., tetrapods, pyramids, etc.)\cite{glotzer2007anisotropy,oleggangACSNano,Escobedo2011,Glotzerscience} or orientationally-specific ``patchy'' interactions.\cite{vega:9938,OgangNL2010,EGNoyaJCP,sciortinonatcom,patchyGKahl,designrulepatchy_glotzer,WangsciencepineNov2012,vissers:164505} What is perhaps surprising is the prediction from computer simulations that simple cubic and diamond crystalline phases can also result from self-assembly of particles with isotropic interactions\cite{B814211B,PhysRevE.74.021404,rechtsmandiamond,YKorig1,finitetempYK,LikosStarPot,GALikos1,MPLikos,fominsc,hertz,groundstatessoftmatterpaper,marcottediamondpaper,edlundjcp2013} including some that are purely repulsive\cite{YKorig1,finitetempYK,LikosStarPot,GALikos1,MPLikos,fominsc,hertz,groundstatessoftmatterpaper,marcottediamondpaper}. These latter systems are of practical interest because, while large quantities of nanoparticles with precise anisotropic interactions can be challenging and expensive to manufacture, those with approximately isotropic interactions (e.g, ligand-coated nanocrystals) can be readily synthesized. Furthermore, since such isotropic pair interactions are ``effective" in nature, they can be tuned by, e.g., the choice of the ligand and the properties of the solvent.\cite{BAK-JPCB-2002,BAK-ss-2002,VlugtTJH} Finally, self-assembly of particles with isotropic interactions may be advantageous from a kinetic perspective when compared to patchy particles, since the energy landscapes of the latter can be particularly rugged.\cite{Energylanscapechargedjanus}

Motivated by an insightful study on statistical mechanical design of isotropic interactions for low-coordinated ground states in two dimensions,\cite{JCP2dmonotonic} we recently introduced an inverse approach\cite{groundstatessoftmatterpaper} for discovering simple pair potentials\footnote{The optimized pair potential \(V(x)\) as a function of nondimensional interparticle separation \(x=r/\sigma\) has the form \unexpanded{
\(
V(x)= \epsilon \{ A x^{-n}+\sum_{j=1}^2\lambda_{i}\left(1-\tanh\left[k_{j}\left(x-\delta_j\right)\right]\right) +
\vshift (x)\} H[\xcut-x]
\).
} Here, \(\epsilon\) and \(\sigma\) are characteristic energy and length scales chosen so that \(V(1)/\epsilon=1\), \(\xcut=2.25\) is the range, \(H\) is the Heaviside step function, \(\vshift (x)=Bx^2+Cx+D\), and constants \{$B$, $C$, $D$\} are chosen so that \(V(\xcut)=V^{\prime}(\xcut)=V^{\prime \prime}(\xcut)=0\).} that exhibit three-dimensional diamond\footnote{For the diamond forming potential \(\vdia\), the optimized parameters are \(A=0.340010893\), \(n=3.39499\), \(\lambda_{\text{1}}=0.732696686\), \(k_{\text{1}}=54.1266\),\(\delta_{\text{1}}=2.77156\), \(\lambda_{\text{2}}=0.605959303\), \(k_{\text{2}}=3.71791\), \(\delta_{\text{2}}=1.08107\), \(B=-0.0375558\), \(C=0.20321\), and \(D=1.17642\)}  or simple cubic \footnote{For the simple cubic forming potential \(\vsc\), the optimized parameters are \(A=0.394620116\), \(n=5.31842\), \(\lambda_{\text{1}}=0.247527755\), \(k_{\text{1}}=58.1066\),\(\delta_{\text{1}}=2.63593\), \(\lambda_{\text{2}}=0.533048319\), \(k_{\text{2}}=4.35419\), \(\delta_{\text{2}}=1.05393\), \(B=-0.0187544\), \(C=0.0971676\), and \(D=0.366054\)} ground states stable {\em over a wide range of densities}. The qualitative characteristics of the interaction potentials we optimized (short-range, isotropic, convex, and repulsive) were inspired by the soft, entropic effective repulsions encountered in many polymer(ligand)-decorated colloidal(nano) particle systems. The optimized interactions we obtained for target ground states with diamond (\(\vdia\)) and simple cubic (\(\vsc\)) structures are only subtly different from one another. For a comparison, see insets (c) and (d) of Figure~\ref{fgr:PressurePhase}. In short, the former has a softer core and a slightly harder shoulder than the latter. As we discuss below, these differences have important implications for the crystalline phases favored at intermediate densities. We refer readers to our previous publication\cite{groundstatessoftmatterpaper} for details on the stochastic optimization routine used to obtain these potentials and the corresponding ground state phase diagrams. A similar pair potential with a diamond ground state was independently discovered by Marcotte, Stillinger, and Torquato using an alternative inverse design strategy with different constraints,\cite{marcottediamondpaper} suggesting that the qualitative form of the optimized interaction is robust.

In this Communication, we use free energy based Monte Carlo (MC) simulation methods to elucidate the effect of temperature (entropy) on the aforementioned diamond forming and simple cubic forming potentials\cite{groundstatessoftmatterpaper} by constructing their equilibrium phase diagrams.  Despite the apparent simplicity of the interparticle interactions, we find that the models display rich, polymorphic phase behavior.  Crystalline phases corresponding to the target simple cubic and diamond ground-state structures feature prominently on their respective phase diagrams as well as a variety of other interesting crystalline phases and multiple reentrant melting transitions.

%Methods:
In order to trace the loci of phase coexistence points in the temperature-pressure-density (\(T-P-\rho\)) space, we first evaluate at least one reference inter-crystal and crystal-fluid coexistence point as follows. We locate crystal melting points along a specific isobar by computing the temperature at which the Gibbs free energy of the crystal equals that of the fluid. We employ Frenkel-Ladd\cite{FLMC} and isothermal-isobaric temperature-expanded ensemble MC simulations\cite{JCP1992,jctcjeffpaper} to determine the temperature dependence of the Gibbs free energy of the crystalline phases and a combination of grand canonical transition matrix MC\cite{JeffGCMC} and isothermal-isobaric temperature expanded ensemble (NPT-TEE) MC\cite{JCP1992,jctcjeffpaper} simulations to determine the Gibbs free energy of the fluid phase as a function of temperature. The inter-crystal coexistence points are evaluated for a single low-\(T\) isotherm (\(T=0.01 \epsilon/\kB\), where \(\kB\) is the Boltzmann constant)--and other isotherms as needed--using a combination of Frenkel-Ladd\cite{FLMC} and density expanded ensemble MC simulations.\cite{jctcjeffpaper}

We trace inter-crystal coexistence curves and fluid-crystal saturation curves by employing a recently introduced technique\cite{jctcjeffpaper} that also uses isothermal-isobaric temperature expanded ensemble (NPT-TEE) MC simulations, wherein the sub-ensembles are differentiated by a shift in temperature and pressure. In this method, one first completes a NPT-TEE simulation with a guess of the temperature-pressure relationship along the saturation line and later corrects this guess by employing histogram reweighting with the relevant probability distributions (e.g., volume, enthalpy) to identify the coexistence point associated with each sub-ensemble sampled. This approach allows us to compute coexistence points over a wide range of temperatures within a single simulation. For tracing inter-crystal coexistence curves, we use NPT-TEE simulations where temperature is the dominant variable and work with the resulting density probability distribution. We then use histogram reweighting to determine the pressure at which the chemical potentials of the two crystal phases are equal at each sub-ensemble temperature. For solving the fluid-crystal coexistence curves, we use another variant of NPT-TEE where pressure is the dominant variable. We utilize the enthalpy probability distribution in this simulation and compute the coexistence temperature at each sub-ensemble pressure by using histogram reweighting. This strategy requires a reference saturation state point and the corresponding Gibbs free energy, which we provide from the single isobar MC simulations explained above. We refer the reader to the earlier publication\cite{jctcjeffpaper} for a detailed description of this approach. The NPT-TEE simulation of the crystal phase is able to sample inter-crystal transitions or crystal-fluid transitions for cases when the specified pressure-temperature sub-ensemble relationship extends to conditions beyond the crystal's stability range. Thus, it often becomes apparent from results of the simulation if another crystal (which might not have been considered in the original free energy calculations) is more stable at a given state point--information that is then used to suggest new free energy calculations to refine the phase diagram. We also completed several quench simulations at different densities, wherein we allow a fluid (initially equilibrated at a high temperature of \(T= \epsilon/\kB\)) to relax at lower temperatures, and we verified that it assembles into the expected crystal.

\begin{figure}
\centering
  \includegraphics[width=8.5cm]{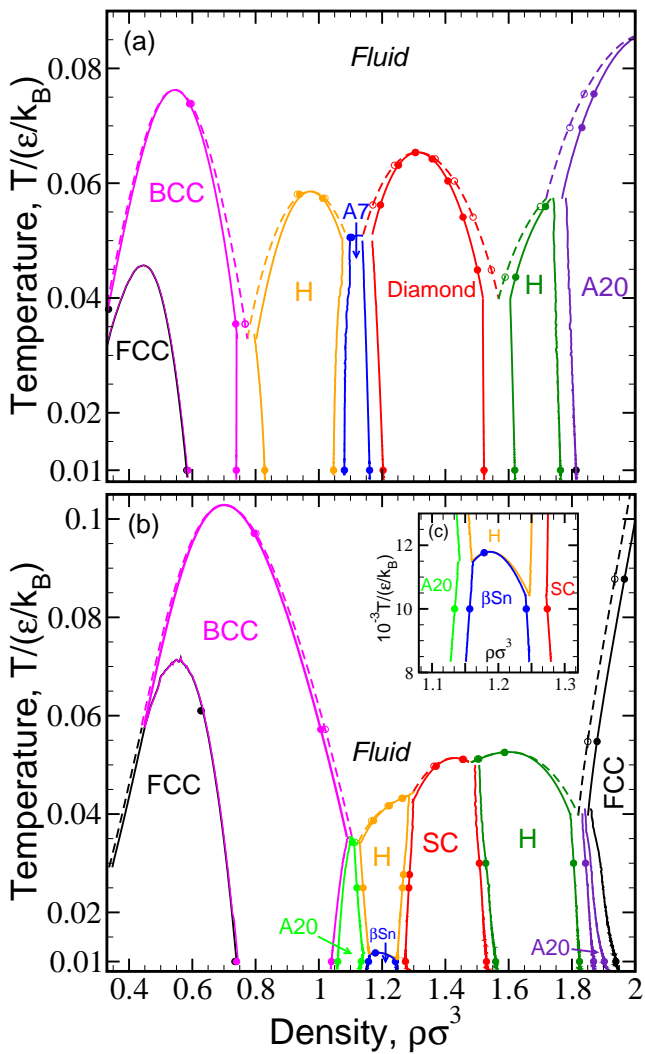}
  \caption{Temperature-density projection of the phase diagram of the (a) diamond forming potential (\(\vdia\)) and (b) simple cubic forming potential (\(\vsc\)). Inset (c) zooms in the region around the \(\beta\)Sn phase. Symbols and curves are as described in Figure~\ref{fgr:PressurePhase}.}
  \label{fgr:DensityPhase}
\end{figure}

\begin{table}[b]
\caption{\label{tab:LatPar}Lattice parameters for the equilibrium crystal phases of the optimized potentials, \(\vdia\) and \(\vsc\) (nomenclature is that of an earlier reference\cite{zerotempsanti}).}
  \setlength{\tabcolsep}{8pt}
\begin{ruledtabular}
\begin{tabular}{ll}
   \noalign{\smallskip}
   \multicolumn{2}{l}{Diamond forming potential model, \(\vdia\)} \\
        \hline
     \noalign{\smallskip}
   Hexagonal\footnote{in the lower density region} & \(c/a:\>[\,0.589-0.734]\) \\
   %\hline
   A7 & \(b/a:\>3.98\), \(u:\>0.317\) \\
   %\hline
   Hexagonal & \(c/a:\>1.4\) \\
   %\hline
   A20 & \(b/a:\>1.73\), \(c/a:\>0.622\),\(\,y:\>0.33\) \\
   \hline
   \noalign{\smallskip}
   \multicolumn{2}{l}{Simple cubic forming potential model, \(\vsc\)} \\
   \hline
   \noalign{\smallskip}
   Hexagonal\(^\text{a}\) & \(c/a:\>[\,0.845-0.89]\) \\
   A20 & \(b/a:\>2.27\), \(c/a:\>1.8\),\(\,y:\>0.9025\) \\
   %\hline
   \(\beta\text{Sn}\) & \(c/a:\>[\,0.66-0.7]\) \\
   %\hline
   Hexagonal & \(c/a:\>[\,0.95-1.0]\) \\
   %\hline
   A20 & \(b/a:\>3.1\), \(c/a:\>1.02\),\(\,y:\>0.867\) \\
   \end{tabular}
  \end{ruledtabular}
  \end{table}

The structures we consider in the free-energy calculations consist of all lattices which occur as ground states for the optimized potentials,\cite{groundstatessoftmatterpaper} as well as other lattices which we found to be slightly less stable than the ground states (by an enthalpy difference of order \(10^{-4}\epsilon\)). This composite list includes face-centered cubic (FCC), body-centered cubic (BCC), simple cubic (SC), diamond (DIA), hexagonal (H), A7, A20, \(\beta\text{Sn}\) and cI16 crystalline phases. We also allow for changes in box shape and in the lattice parameters during the simulations. The MC simulations typically contained \(700-2000\) particles, the actual number depending on the lattice-type. We simulated larger sized simulations with \(\geq\,3000\) particles and did not observe significant differences in the free energies.

%Discussion
We present the computed projections of the fluid-crystal and inter-crystal phase coexistence curves in the \(P-T\) and \(T-\rho\) planes in Figures~\ref{fgr:PressurePhase} and \ref{fgr:DensityPhase}, respectively.  First, we observe that the target phases (diamond for \(\vdia\) and simple cubic for \(\vsc\)) feature prominently on the corresponding phase diagrams. This is notable because such phases rarely occur in systems with isotropic interactions, and--when they have appeared in model systems--they generally display a narrow range of thermodynamic stability.\cite{groundstatessoftmatterpaper} In this case, however, the potentials were explicitly designed to exhibit the target structures in the ground state (at \(T=0\)) over a wide range of densities.\cite{groundstatessoftmatterpaper} What our new calculations show is that the crystalline phases corresponding to the ground-state optimization targets also exhibit good thermal stability (i.e., comparable to other stable lattices on the phase diagram). This is encouraging because it suggests that focusing exclusively on the ground-state structures in the optimization phase of the pair-potential design can be an effective strategy, even if the ultimate goal is stability at finite temperature. We further note that several structures for which the potentials were not optimized (e.g., \(\beta\)-Sn, rhombohedral, and body-centered orthorhombic lattices), but that nonetheless appear in the ground-state phase diagram,\cite{groundstatessoftmatterpaper} display poor thermal stability with only \(\beta\)-Sn surviving to the lowest temperatures explored here. The ground-state optimization of the potentials for the target lattices against a large pool of competitive structures across a wide density range may well have had the secondary effect of increasing the relative thermal stability of the target phases. We may gain more insight into this hypothesis by constructing the equilibrium phase diagrams of interaction potentials designed using different inverse methods\cite{marcottediamondpaper,rechtsmandiamond,edlundjcp2013}.

In terms of the qualitative features of the phase diagrams in Figures~\ref{fgr:PressurePhase} and \ref{fgr:DensityPhase}, we find that there are clear similarities between the two models, which is expected given the corresponding similarities in the pair interactions. Both models feature reentrant melting transitions and many of the same phases including FCC, BCC, low- and high-density hexagonal, and A20 crystals. The fluid-FCC-BCC transitions appearing at low particle densities are similar to those seen in other model soft materials like Hertzian spheres~\cite{hertz}, the Gaussian core model~\cite{GCMmodelPRE}, and a model for ionic microgels~\cite{phaseionicmicrogel}. The most striking difference in the phase diagrams of the two models is the presence of the associated target phase--either diamond or simple cubic. Thus, at least for the short-range, soft repulsive potentials considered here, it appears that the potential optimization has a relatively localized effect, having a significant qualitative impact on the phases that appear at intermediate densities.    

%Conclusion 
To summarize, we have used advanced free energy based MC simulations to precisely compute the phase diagrams for two potential models \(\vdia\) and \(\vsc\), which have previously been designed\cite{groundstatessoftmatterpaper} to exhibit stable diamond or simple cubic lattice ground states over a wide range of densities. The approach traces out crystal-crystal and fluid-crystal saturation curves within one molecular simulation rather than simulating pairs of phases at several isotherms or isobars to determine the coexistence behavior. We find that the target phases in the models show good thermal stability relative to other potentially competitive crystalline forms, suggesting that our ground-state optimization may also have contributed to the thermal stability of the simple cubic and diamond phases. Despite the simplicity of the model, the overall phase behavior is rich, exhibiting reentrant melting transitions, non-Bravais A20, A7 and \(\beta\text{Sn}\) crystals, as well as Bravais FCC, BCC and hexagonal crystalline phases.

%Closing arguments
Although qualitative features of the optimized models studied here are motivated by the form of effective repulsive forces observed in colloidal and nanoparticle systems, a critical test will be to carry out similar computations on potential models which exhibit a more direct relationship between the model variables and experimentally-controllable parameters such as nano-(colloidal) particle type, ligand length, and solvent density. We are currently investigating such a system with experimental collaborators, and we will report on our findings in a future publication.

%more realistic, experimentally realizable effective pair potentials.

%Acknowledgements
T.M.T. acknowledges support of the Welch Foundation (F-1696) and the National Science Foundation (CBET-1065357). J.R.E. acknowledges support of the National Science Foundation (CHE-1012356).  We also acknowledge the Texas Advanced Computing Center (TACC) at The University of Texas at Austin and the Center for Computational Research at the University at Buffalo for providing HPC resources that have contributed to the research results reported within this paper.

\balance

%\nocite{*}
\bibliography{PhaseDiagram_bib}% Produces the bibliography via BibTeX.

\end{document}